\documentclass{article}
\begin{document}
\begin{center}
\begin{title}
 ~\Large{Particle interactions, masses and the symmetry breaking}\\
\end{title}

~

\begin{author}
~Fizuli Mamedov
\end{author}

~

 \textit{~Institute of Physics, Baku, Azerbaijan}
\end{center}

\begin{abstract}
A possible original $SU(2)_{L} \times SU(2)_{R}$ symmetry of the
elementary particles and the mechanism of its breaking is discussed.
It is concluded that it is the broken symmetry states of the
particles which induce the interactions among the particles.
\end{abstract}

We briefly review the unbroken $~SU(2)_{L} \times U(1)_{Y}~$
symmetry part of the Standard Model of electroweak interactions
first for the sake of clarity of the further discussions. According
to this theory, the original Lagrangian of the (massless) particles
possess both $SU(2)$ and $U(1)_{Y}$ symmetries for the left - handed
fermion doublets. The right - handed fermions have $U(1)_{Y}$
symmetry only. The triplet of the massless vector gauge bosons
associated with the $SU(2)_{L}$ is $\textbf{W}_{\mu}$ and the
massless vector gauge boson associated with the $U(1)_{Y}$ is
$B_{\mu}$. The Lagrangian for the unbroken $SU_{L}\times U(1)_{Y}$
symmetry is

\begin{equation}\label{Lagr}
 L = -\frac{1}{4}\textbf{W}^{\mu\nu}\textbf{W}_{\mu\nu} -
 \frac{1}{4}B^{\mu\nu}B_{\mu\nu} +\bar{\Psi}i\gamma^{\mu}D_{\mu}\Psi
\end{equation}

\noindent with $\Psi$ denoting the fermionic fields.

\noindent The covariant derivative $~D_{\mu}~=~\partial_{\mu} ~+ ~i
g$ $\textbf{W}_{\mu}\dot{}\textbf{T}~+ ~\frac{1}{2}i g\prime$
$B_{\mu}~Y$. $\textbf{T}$ denotes the Pauli matrices and Y is the
hypercharge of the particles. The second term in $D_{\mu}$ act only
on the left - handed fermions. The Lagrangian is invariant under the
infinitesimal local gauge transformations for $SU(2)_{L}$ and
$U(1)_{Y}$:

~~

$SU(2)_{L} $

\begin{center}
 $\Psi_{L} \rightarrow [1 - i g T\dot{}\alpha(x)]\Psi_{L}$
 $ ~~~~~~~~~~~~~~~~~~~~~~~~~~~\Psi_{R} \rightarrow \Psi_{R}$
\end{center}
\begin{equation}\label{su2lefttr}
W_{\mu}\rightarrow W_{\mu}+\partial_{\mu}\alpha(x)+g \alpha(x)\times
W_{\mu} ~~~~~~~~~~~~B_{ \mu} \rightarrow B_{\mu}
\end{equation}

~

 $U(1)_{Y}$

\begin{center}
$\Psi_{L} \rightarrow [1 - i\frac{1}{2}g\prime]Y\beta(x)\Psi_{L}
 ~~~~~~~~~~~~~~~~~~~~~~~~~ \Psi_{R} \rightarrow [1 -i \frac{1}{2}g\prime]Y\beta
(x)\Psi_{R}$
\end{center}
\begin{equation}\label{u1tr}
W_{\mu} \rightarrow W_{\mu}
~~~~~~~~~~~~~~~~~~~~~~~~~~~~~~~~~~~~~~~~~~~~~ B_{\mu} \rightarrow
B_{\mu}+\partial_{\mu}\beta(x)
\end{equation}

The following definitions allow to write the interaction Lagrangian
of the fermions and gauge bosons in a compact form:
\textbf{$~W\dot{}T$} ~$=~ W^{+}\dot{}T^{+} ~+ ~W^{-}\dot{}T^{-}~ +~
W_{3}\dot{}T_{3}~$, where $~W_{\mu}^{\pm}~ =~
\frac{1}{\sqrt{2}}(W_{1\mu} ~ \mp~ i W_{2\mu}) ~$ and $~T^{\pm} ~=~
\frac{1}{\sqrt{2}}( T_{1} ~ \pm ~i T_{2})~$. The electromagnetic
field $A_{\mu}$ should be contained in the $~i g W_{ 3\mu }T_{3}~ +~
i g\prime \frac{1}{2} B_{\mu}Y~$, in the neutral term of the
covariant derivative. By defining $~A ~= ~W_{3}sin\theta_{w}~ + ~B
cos\theta{w}~ $ and $~Z~=~W_{3}cos\theta_{w}~ - ~B sin\theta{w} ~$,
this term can be written as

\begin{center}
$~i g W_{3}T_{3}~ + ~i g\prime \frac{1}{2}B Y ~= ~i \textit{A}[g sin
\theta_{w}T_{3}~ +~ g\prime cos\theta_{w} \frac{1}{2} Y] ~+ ~i Z[g
cos \theta_{w}T_{3}~ -~ g\prime sin\theta_{w} \frac{1}{2}Y]~$
\end{center}

\noindent which yields the quasi - Gell - Mann -Nishijima formula
for the electric charge: $~i e Q = i e(T_{3} + \frac{1}{2}Y)$~. For
the left - handed fermions $T_{3} = \pm \frac{1}{2}$ and for the
right - handed fermions $T_{3} = 0$. Also $~g =
e/sin\theta{w},~~g\prime = e/cos\theta{w},~~g_{Z} = e/(sin\theta_{w}
cos\theta_{w}), ~~x_{w} = sin^{2}\theta_{w}$.

In terms of the $W_{\mu}^{\pm}$, $A_{\mu}$ and $Z_{\mu}$ fields the
fermion - gauge boson interaction Lagrangian is written as

\begin{equation}\label{intLagr}
 -L^{\prime}~ = ~e J_{em}^{\mu}A_{\mu}~ + ~\frac{g}{\sqrt{2}}\large{(}J_{L}^{+\mu}W_{\mu}^{+}~
 +~ J_{L}^{-\mu}W_{\mu}^{-}\large{)}~+~ g_{Z}J_{Z}^{\mu}Z_{\mu}
\end{equation}

\noindent Here
$~J_{L}^{\pm\mu}~=~\sqrt{2}\bar{\Psi}\gamma^{\mu}T_{L}^{\pm}\Psi,~~J_{Z}^{\mu}$
$~=~ \bar{\Psi}\gamma^{\mu}[T_{3L}~-~$
$x_{w}Q]\Psi~~$and$~~J_{em}^{\mu}~=~\bar{\Psi}\gamma^{\mu}Q\Psi~~$
are the fermion - gauge field interaction currents ($T_{L}$ acts on
the left - handed fermions only). The Standard Model scenario
spontaneous symmetry breaking does not affect this form of the
interaction currents.

 It is possible that the symmetry breaking occurs because of the
fusion and condensation  of the certain type of the fermions. In
such a picture only massless fermionic fields exist at the initial,
highest symmetry stage. Left - handed and right - handed fermions
manifest themselves as two separate irreducible representations of
the most fundamental field. Besides the space-time spin $s$,
fermions also have the electroweak spin\footnote{We use the usual
weak isospinor notation, $I_{W}$ for the electroweak spin in this
work} $I_{W} = \frac{1}{2}$ for both left -handed and right -handed
fermions. The fused and condensed states of the left - handed
fermions can be separated into electroweak spin triplet $I_{W}=1$
and singlet $I_{W}=0$ states.

\begin{center}
$|s_{L}=\frac{1}{2}>|I_{W}=\frac{1}{2}, m_{W}=\pm \frac{1}{2}>
~\times ~|s_{L}=\frac{1}{2}>|I_{W}=\frac{1}{2}, m_{W}=\pm
\frac{1}{2}> ~=~$
\end{center}

\begin{equation}\label{fusionl}
|s_{L}= 1>(|I_{W}=1> ~+ ~|I_{W}= 0>)
\end{equation}

The electroweak triplet states are represented by the triplet of the
vector gauge boson fields\footnote{It is the space - time parts of
these fields that are treated as classical fields at first and after
the second quantization become quantum fields}
$~(\textbf{W}^{\mu})_{L}~$ and the singlet state is represented by
the vector boson field $~(B^{\mu})_{L}~$. The similar fusion and
condensation occurs for the right - handed fermions.

\begin{center}
$|s_{R}=\frac{1}{2}>|I_{W}=\frac{1}{2}, m_{W}=\pm \frac{1}{2}>
~\times~ |s_{R}= \frac{1}{2}>|I_{W}=\frac{1}{2}, m_{W}=\pm
\frac{1}{2}> ~=~$
\end{center}

\begin{equation}\label{fusionr}
 |s_{R}=1>(|I_{W}=1> ~+~ |I_{W}=0>)
\end{equation}

The states with $s_{L} =1$, $~(\textbf{W}^{\mu})_{L}~$ and
$~(B^{\mu})_{L}~$ couple to the left-handed fermions only and the
states with $s_{R} =1$, $~(\textbf{W}^{\mu})_{R}~$ and
$~(B^{\mu})_{R}~$ couple to the right - handed fermions only.

The states $~\frac{1}{\sqrt{2}}((W^{\mu}_{1})_{L} +
i(W^{\mu}_{2})_{L})~$, $\frac{1}{\sqrt{2}}((W^{\mu}_{1})_{L} -
i(W^{\mu}_{2})_{L})~$ and $~(W^{\mu}_{3})_{L}~$ correspond to the
three generators of the irreducible representation of the the
$~SU(2)_{L}~$ ($I_{W}=1$) in the spherical coordinates. Therefore
the fields

\noindent $~\frac{1}{\sqrt{2}}((W^{\mu}_{1})_{L} ~\pm~$
$i(W^{\mu}_{2})_{L})~$ should be associated with the $I_{W} = 1,~
m_{W}= \pm 1$ states, and the field $W^{\mu}_{3}$ with the $I_{W} =
1,~ m_{W} = 0 $ state. Thus all the particles (including the
fermions) correspond to the combinations of the fields with the
definite value of $m_{W}$ (the role of the $m_{W}$ is played by
$T_{3}$ in the context of the $~ SU(2)_{L} \times U(1)_{Y} ~$
symmetry, discussed earlier in this work). Since both
$~(W^{\mu}_{3})_{L}~$ and $~(B^{\mu})_{L}~$ have $m_{W} = 0$, their
mixtures will also have the definite value for the $m_{W}$, $m_{W}
=0$, and can exist as separate particles. The (massive) $Z$
boson\footnote{We will often drop the indices $L$ and $R$ when their
presence is not important} and the (massless) photon are indeed two
such particles. Nature separates the 'steam' part of the $m_{W} = 0$
mixed states, the massless vector boson from their condensate part,
the massive vector boson. Interestingly, the fields with $m_{W} =0$,
the $Z$ bosons and photons carry the interactions between the same
type of fermions, the particles of the same handedness, $m_{W}$ and
generation. On the other hand, fields with $|m_{W}| = 1$, the
$W^{\pm}_{\mu}$, couple fermions with the different $m_{W}$ only and
can couple particles of the different generations.

Let us assume that the $\textbf{W}^{\mu}$ gain the mass $M_{W}$ and
the $B^{\mu}$ gain the mass $M_{B}$ as a result of the symmetry
breaking. It is natural to assume that, due to the universality of
the condensation and couplings,

\begin{equation}\label{massrel}
M_{W}/g ~=~ M_{B}/g^{\prime}
\end{equation}

This equation implies that the masses gained by the $\textbf{W}$ and
$B$ due to the symmetry breaking are proportional to their
respective coupling strengths, $M_{W}~=~k g~$ and $M_{B}
~=~k^{\prime} g^{\prime}$, with $k~=~k^{\prime}$.
Eq.($\ref{massrel}$) leads to the usual definition of $Z_{\mu}$ and
to the well - known $M_{W}$ - $M_{Z}$ mass relation:

\begin{center}
$M_{W}^{2}W_{1}^{2}~+~M_{W}^{2}W_{2}^{2}~+~M_{W}^{2}W_{3}^{2}~+~
M_{B}^{2}B^{2} =$
\end{center}

\begin{center}
$M_{W}^{2}W^{+}W^{-}~+~(M_{W}^{2}/g^{2})(g^{2}W_{3}^{2}~+~
(g^{\prime})^{2}B^{2})~=~$
\end{center}

\begin{center}
$M_{W}^{2}W^{+}W^{-}~+~(M_{W}^{2}/g^{2})(gW_{3}~-~g^{\prime}B)^{2}
~+~4(M_{W}^{2}/g^{2})gg^{\prime}W_{3}B =$
\end{center}

\begin{center}
$M_{W}^{2}W^{+}W^{-}~+~M_{Z}^{2}Z^{2} ~$
\end{center}

\noindent with
$M_{Z}~=~M_{W}\sqrt{g^{2}~+(~g^{\prime})^{2}}/g~=~M_{W}/cos\theta_{W}~~$
and

\noindent$~Z~=~(gW_{3}~-~g^{\prime}B)/\sqrt{g^{2}~+~(g^{\prime})^{2}}$
$~=W_{3}cos\theta_{W}~-~Bsin\theta_{W}~$. The term
$~4(M_{W}^{2}/g^{2})gg^{\prime}W_{3}B~$ in this equation disappears
because $~W_{3}~$ and $~B~$ are orthogonal: these states belong to
the different multiplets of the electroweak spin $I_{W}$.

The known values for the electroweak spin and hypercharge of the
particles leads to the following possible scenario of the symmetry
breaking process within this model. At first only $B_{\mu}$ fields
couple to the fermions. The $~(B^{\mu})_{L}~$ to the left - handed
fermions and the $~(B^{\mu})_{R}~$ to the right - handed fermions
only. The fermion doublets have $SU(2)_{L} \times SU(2)_{R}$ global
symmetry at this stage with the electroweak quantum numbers:

~~\\

 $\nu_{L}~~ m_{W} = \frac{1}{2},  ~~ Y =  -1$
 $~~~~~~~~~~~~e_{L} ~~ m_{W} = -\frac{1}{2}, ~ ~Y = -1$

 ~~\\

$u_{L} ~~ m_{W} = \frac{1}{2}, ~ ~ Y =  \frac{1}{3}$
 $~~~~~~~~~~~~~~d_{L}~~  m_{W} = -\frac{1}{2}, ~~ Y = \frac{1}{3}$

~~\\

 $\nu_{R}~~ m_{W} = \frac{1}{2},  ~~Y =  -1 $
 $~~~~~~~~~~~~e_{R}~~  m_{W} = -\frac{1}{2}, ~~ Y = -1$

~~\\

 $u_{R}~~  m_{W} = \frac{1}{2},  ~~ Y =  \frac{1}{3}$
 $~~~~~~~~~~~~~~d_{R}~~  m_{W} = -\frac{1}{2}, ~~ Y =   \frac{1}{3}$

~~\\

The $~SU(2)_{L} \times SU(2)_{R}~$ global symmetry breaking occurs
when $~(B^{\mu})_{L}~$ also couples to the right - handed fermions.
Simultaneously, the direct left - right handed fermion couplings,
the masses of the fermions appear. $~(B^{\mu})_{R}~$ does not couple
to the left - handed fermions: in that case the original
$~SU(2)_{L}\times SU(2)_{R}~$ global symmetry would be preserved.
Typically the emergence of the interactions (couplings) is
accompanied with the movement of the system to the less symmetrical
state. For example, the inclusion of the two - particle interactions
between the atoms replaces the global $SO(3)$ symmetry by the global
$U(1)$ symmetry. Besides, the $~(B^{\mu})_{L}~$ couples to the upper
members of the right - handed doublets with the strength $Y=1$ and
to the lower members of these doublets with the strength $Y=-1$ (the
$~(B^{\mu})_{L}$ couples 'color - blindly', with $|Y|=1$, to the
quarks of the different irreducible representation of the most
fundamental field, to the $u_{R}$ and $d_{R}$). Once again, the same
$~(B^{\mu})_{L}~$ couplings to these particles would preserve the
right - handed doublets and hence the global $SU(2)_{L} \times
SU(2)_{R}$ symmetry. As a result, the well - known values of the
electroweak quantum numbers of the particles are produced:

 ~~\\

 $\nu_{L}~~ m_{W} = \frac{1}{2},  ~~ Y =  -1$
 $~~~~~~~~~~~~e_{L} ~~ m_{W} = -\frac{1}{2}, ~ ~Y = -1$

 ~~\\

$u_{L} ~~ m_{W} = \frac{1}{2}, ~ ~ Y =  \frac{1}{3}$
 $~~~~~~~~~~~~~~d_{L}~~  m_{W} = -\frac{1}{2}, ~~ Y =   \frac{1}{3}$

~~\\

 $\nu_{R}~~ m_{W} = 0,  ~~Y =  0 ~~$
 $~~~~~~~~~~~~e_{R}~~  m_{W} = 0, ~~ Y = -2$

~~\\

 $u_{R}~~  m_{W} = 0,  ~~ Y =  \frac{4}{3}$
 $~~~~~~~~~~~~~~d_{R}~~  m_{W} = 0, ~~ Y =   -\frac{2}{3}$

~~\\

The $~(\textbf{W}^{\mu})_{L}~$ triplet of the vector bosons'
coupling to the left - handed fermions  also occurs at this stage.
From Eq.(\ref{massrel})  we see that indeed a non-zero value of $g$
produces a non-zero value for $M_{W}$, once the $B^{\mu}$ gains its
mass at the stage of its own coupling to the fermions. With no right
- handed fermion doublets around, $~(W^{\mu})_{R}~$ does not couple
to the fermions. Also, the coupling of the $~(W^{\mu})_{R}~$ to the
fermions would produce a local $~SU(2)_{L} \times SU(2)_{R}~$
symmetry, not a typical consequence of the appearance of the new
interactions. Thus the emergence of the new types of interactions
among the particles leads to the disappearance of the global
$~SU(2)_{L} \times SU(2)_{R}~$ symmetry, along the way also breaking
the space left - right symmetry. The fusion of the fermions and
generation of the $B^{\mu}$ mass also breaks the symmetry, the
original symmetry of the most fundamental field, the irreducible
representations of which are the non - interacting massless
fermions. Thus we conclude that the interactions (couplings) among
the particles are the manifestations of the symmetry breaking. In
other words, it is the broken symmetry states that induce the motion
of the particles along the curved paths and the mass, as in the case
of the gravitational field, is the measure of this curvedness. It is
interesting to notice that the 'unbroken part' of the $W_{3}^{\mu} -
B^{\mu}$ mixture, the photons do not interact with the neutrinos,
the particles  which avoid at least the mass generation consequence
of the symmetry breaking.

It is the $s_{L}=1$ state with $I_{W}=0$, $~(B^{\mu})_{L}~$, that
couples to the right - handed fermions after the $~SU(2)_{L}\times
SU(2)_{R}~$ symmetry is broken. Therefore the left - handed fermions
and the right - handed fermions might have two distinct electroweak
spin quantum numbers, $(I_{W})_{L}=\frac{1}{2}$ and
$(I_{W})_{R}=\frac{1}{2}$, respectively.

The fermion - vector boson interaction Lagrangian in the form of
Eq.(\ref{intLagr}) appears only after the breaking of the global
$~SU(2)_{L} \times SU(2)_{R}~$ symmetry, after its replacement of by
the $U(1)$ electromagnetic gauge symmetry. As for the $SU(2)_{L}$
gauge symmetry transformations, Eq.(\ref{su2lefttr}) and, perhaps,
$U(1)_{Y}$ symmetry transformations, Eq.(\ref{u1tr}), they are never
realized for the electroweak interactions Lagrangian  within this
model.

\begin{center}
    \textbf{References}
\end{center}

\begin{enumerate}
  \item S. L. Glashow, \textit{Nuclear Physics} \textbf{22}
  (1961), 579
  \item S. Weinberg, \textit{Physical Review Letters}, \textbf{19},
  1264
  \item A. Salam, \textit{Svartholm, N, Proceedings of the
  Eighth Nobel Symposium, Almqvist \& Wiksell, Stockholm} (1968)
  \item V. Barger, R. Philips, \textit{Collider Physics, Addison
  - Wesley Publishing Company, Inc}, (1997)
\end{enumerate}

\end{document}